# Evolution of Ferromagnetism and Electrical Resistivity in Sb-Doped $Cr_4PtGa_{17}$

*Chaoguo Wang,*[a] *Gina Angelo,*[a] *Jeremy G. Philbrick,*[b] *Tai Kong*[b,c] *and Xin Gui*[a]*

[a] Department of Chemistry, University of Pittsburgh, Pittsburgh, PA, 15260, USA
[b] Department of Physics, University of Arizona, Tucson, AZ, 85721, USA
[c] Department of Chemistry and Biochemistry, University of Arizona, Tucson, AZ, 85721, USA
Address correspondence to E-mail: xig75@pitt.edu

## *Abstract*

We describe the doping effects on a metallic breathing pyrochlore compound, $Cr_4PtGa_{17}$. Upon doping with Sb, i.e., $Cr_4Pt(Ga_{1-x}Sb_x)_{17}$, it was found that a selective doping on one of the seven Ga sites occurs. With increasing dopant level, the ferromagnetism in the parent compound is gradually suppressed, along with a decrease in fitted Curie-Weiss temperature (from 61 (1) K to -1.8 (1) K) and effective moment (from ~2.26 $\mu_B$/f.u. to ~0.68 $\mu_B$/f.u.). Low-temperature heat capacity measurements confirm the absence of magnetic ordering above 0.4 K for the three most doped samples. Meanwhile, electrical resistivity measurements display a metal-semiconductor transition with increasing Sb contents, which is attributed to an increase of Fermi energy based on the calculations of electronic band structure and density of states. Moreover, we have speculated that the ferromagnetism in $Cr_4PtGa_{17}$ is governed by itinerant electrons according to our observations. This study of Sb-doping effect on $Cr_4PtGa_{17}$ provides deeper understanding of magnetism of this system and possibilities for future modifications.

## *Introduction*

Frustrated magnetism commonly originates from competition between various types of exchange interaction ($J_{ex}$) as well as thermal energy ($k_BT$). The study of this phenomenon has been of great importance and fundamental research interest in the condensed-matter physics community for decades. Many intriguing quantum states have been proposed/observed, such as quantum spin liquid,[1–4] spin glass[5,6] and quantum spin ice[7,8]. In the meantime, discovering new frustrated magnets is a long-standing goal for material chemists. Frustrated magnetism is commonly accompanied by specific structural motifs of magnetic species, such as triangular,[9–13] square net,[14–16] Kagome[17,18] and pyrochlore[19–21] lattices. From a material design perspective, new findings of materials with these structural motifs and modification of the magnetic interactions in them are a straightforward route to approach intriguing frustrated magnets.

As one of the promising motifs for frustrated magnetism, materials with pyrochlore lattice, i.e., corner-sharing tetrahedral framework, have been heavily studied.[19–22] However, as a derivative of pyrochlore lattice, materials with breathing pyrochlore lattice are rather underdeveloped in literature. A breathing pyrochlore lattice has alternating bigger and smaller corner-sharing tetrahedral framework. Only a few examples of breathing pyrochlore materials have been reported, including $LiGaCr_4O_8$ and its derivatives,[23,24] $Ba_3Yb_2Zn_5O_{11}$,[25,26] and $Cr_4PtGa_{17}$ (CPG)[27]. $LiGaCr_4O_8$ and its derivatives were found to be antiferromagnetically ordered at low temperatures[23] while no magnetic ordering was detected in $Ba_3Yb_2Zn_5O_{11}$ down to 70 mK, making it a promising material for spin liquid.[25] Moreover, ferromagnetism with unusually small magnetic moments and half-metallicity,[27] nearly isotropic magnetic behavior[28] and magnetic clusters[29] were found for CPG.

Considering the breathing pyrochlore lattice in CPG, herein, we performed Sb doping on the Ga site of CPG, i.e., $Cr_4Pt(Ga_{1-x}Sb_x)_{17}$ (CPGS) where $0 \leq x \leq 3.29$ (5)%, to investigate the evolution of magnetic interactions and to interpret the magnetism in this system. With the highly geometrically frustrated breathing pyrochlore lattice of Cr, we aimed to investigate how magnetism is affected by Sb doping, as well as if frustrated magnetism can be induced CPGS. We have found that all Ga atoms that participate in the coordination with Cr and Pt atoms are resistant to Sb doping, while only one Ga site that does not bond with Cr and Pt atoms can mix with Sb. With Sb doping, the ferromagnetism in CPG is gradually suppressed while the electrical resistivity starts to exhibit semiconducting behavior. Heat capacity measurements confirmed the absence of magnetic ordering above 0.4 K for the three most doped CPGS samples. No evidence of magnetic frustration was observed. Assisted by the theoretical calculations on the electronic structure, we speculate that the magnetism in CPG is likely to be governed by itinerant electrons, similar to that of $La_5Co_2Ge_3$.[30] This study deepens the interpretation of magnetism in CPG and provides hints for further modifications on the magnetic interactions in this system.

## *Experimental Details*

**Single Crystal Growth:** Crystals of CPGS were grown using Ga self flux.[27] Elemental Cr (99.94%, ~ 200 mesh, Thermo Scientific), Pt (99.98%, -200+400 mesh, Thermo Scientific) and Sb (99.999%, ~200 mesh, Thermo Scientific) were ground and placed in an alumina crucible with the stoichiometry shown in Table S1 in the Supporting Information (SI). Ga pieces (99.999%, ingot, Thermo Scientific) were subsequently added to the crucibles which were sealed in evacuated quartz tubes. Quartz wool and glass pieces were added above the crucibles to facilitate the separation between Ga flux and obtained crystals. Heavier doping of Sb was tested but the results were indistinguishable from the highest Sb level shown in Table S1 in the SI. Excess Sb in higher Sb batches formed $PtSb_2$ crystals, which reduced the yield of CPGS. The samples were heated to 1050 $^oC$ and held 48 hours before cooling to 450 $^oC$ at a rate of 3 $^oC$ per hour, where the excess Ga flux was centrifuged out. The obtained crystals are not air- and moisture-sensitive and can be washed by hydrochloric acid (1 M) at 80 $^oC$ to remove any Ga residue on the surface.

**Single-Crystal and Powder X-Ray Diffraction:** For each composition, multiple CPGS crystals (~40×40×15 $\mu m^3$) were studied by single crystal X-ray diffraction (XRD) to determine the crystal structure and the doping level. The structure, consistent among all crystals, was determined using a Bruker D8 QUEST ECO diffractometer equipped with APEX4 software and Mo radiation ($\lambda_{K\alpha}$= 0.71073 Å) at ~293 K. The crystals were mounted on a Kapton loop protected by glycerol. Data acquisition was made *via* the Bruker SMART software with corrections for Lorentz and polarization effects included. A numerical absorption correction based on crystal-face-indexing was applied using *XPREP*. The direct method and full-matrix least-squares on $F^2$ procedure within the SHELXTL package were employed to solve the crystal structure.[31,32] Powder XRD patterns, obtained with a Bruker D2 PHASER with Cu Kα radiation and a LynxEye-XE detector, were employed to determine the phase purity of the crystals whose properties were studied. The patterns were fitted by the Rietveld method in Fullprof[33] using the crystal structure obtained from the single crystal data.

**Physical Property Measurement:** The DC magnetization was measured in a Quantum Design physical property measurement system (PPMS) dynacool (1.8- 300 K, 0- 9 T) equipped with an ACMS II function from 2 to 300 K under various applied magnetic fields. All magnetic measurements were carried out on crushed single crystals because large single crystals were not available for higher Sb concentration. Thus, crystals could not be uniformly oriented. Field-dependent magnetization data was collected at multiple temperatures with applied magnetic fields ranging from -9 T to 9 T. The resistivity measurements were carried out in the PPMS using the four-probe method between 1.8 K to 300 K. Platinum wires were attached to the samples by silver epoxy to ensure ohmic contact. Heat capacity was measured using a standard relaxation method in the PPMS, with a $^3$He function for data below 1.8 K.

**Electronic Structure Calculations:** The electronic band structure and electronic density of states (DOS) of $Cr_4PtGa_{17}$ and $Cr_4PtGa_{16}Sb$, i.e., $Cr_4Pt(Ga_{0.9411}Sb_{0.0589})_{17}$ were calculated using the WIEN2k program, which employs the full-potential linearized augmented plane wave method (FP-LAPW) with local orbitals implemented.[34] The electron exchange-correlation potential used to treat the electron correlation was the generalized gradient approximation.[35] The conjugate gradient algorithm was applied, and the energy cutoff was set at 500 eV. For $Cr_4PtGa_{16}Sb$, only $Ga_Z$ site is replaced by Sb atom, see Results and Discussion section. Reciprocal space integrations were completed over an 8×8×2 Monkhorst-Pack *k*-point mesh for both models.[36] Spin-orbit coupling (SOC) effect were only applied for the Pt and Sb atoms. Spin-polarization (ferromagnetism with the moment oriented in the (010) direction) was only employed for the Cr atom in $Cr_4PtGa_{17}$. The structural lattice parameters obtained experimentally were used in all calculations. With these settings, the calculated total energy converged to less than 0.1 meV per atom.

## *Results and Discussion*

**Crystal structure and change in lattice parameters upon Sb doping in CPG:** The crystal structures of all $Cr_4Pt(Ga_{1-x}Sb_x)_{17}$ (CPGS) crystals are identical with undoped CPG, which crystallizes in the noncentrosymmetric space group *R* 3*m* (No. 160). The crystallographic data, refined anisotropic displacement parameters and equivalent isotropic thermal displacement parameters of all CPGS crystals are summarized in Tables S2-S4 in the SI. Figure 1a shows the crystal structure of undoped CPG with Pt@$Ga_6$ octahedra and Cr@$Ga_9$ triply capped trigonal prisms. It was reported to be a half-Heusler type (XYZ) material and the formula can be expressed as $(PtGa_2)(Cr_4Ga_{14})Ga$ (X = $PtGa_2$, Y = $Cr_4Ga_{14}$, Z = Ga).[27] The "isolated" Z = Ga site, i.e., $Ga_Z$ (Wyckoff position: 3*a*) site, is highlighted in red in Figure 1b, which is not bonded to either Cr or Pt atoms. Surprisingly, Sb is selectively doped on the $Ga_Z$ site, although six other Ga sites are available in CPG. The results are proved by single-crystal XRD refinement. When additional Ga sites were assigned disordering with Sb, the refinement became unstable, or the refinement results worsened. The lattice parameters increase monotonically with higher Sb doping, as can be seen in Figure 1c, which is consistent with the results after Rietveld refinements on powder XRD patterns shown in Figure 2a, as discussed later. The increase in lattice parameters is also consistent with Sb atoms are larger than Ga atoms. The highest observed overall concentration of Sb is 3.29 (5)%, which corresponds to a site occupancy of 56 (1)% on the $Ga_Z$ site. The lattice parameter *a* increases ~0.5% from 8.0256 (1) Å to 8.0638 (1) Å, while *c* also rises ~0.5% from 19.6539 (4) Å to 19.7498 (3) Å.

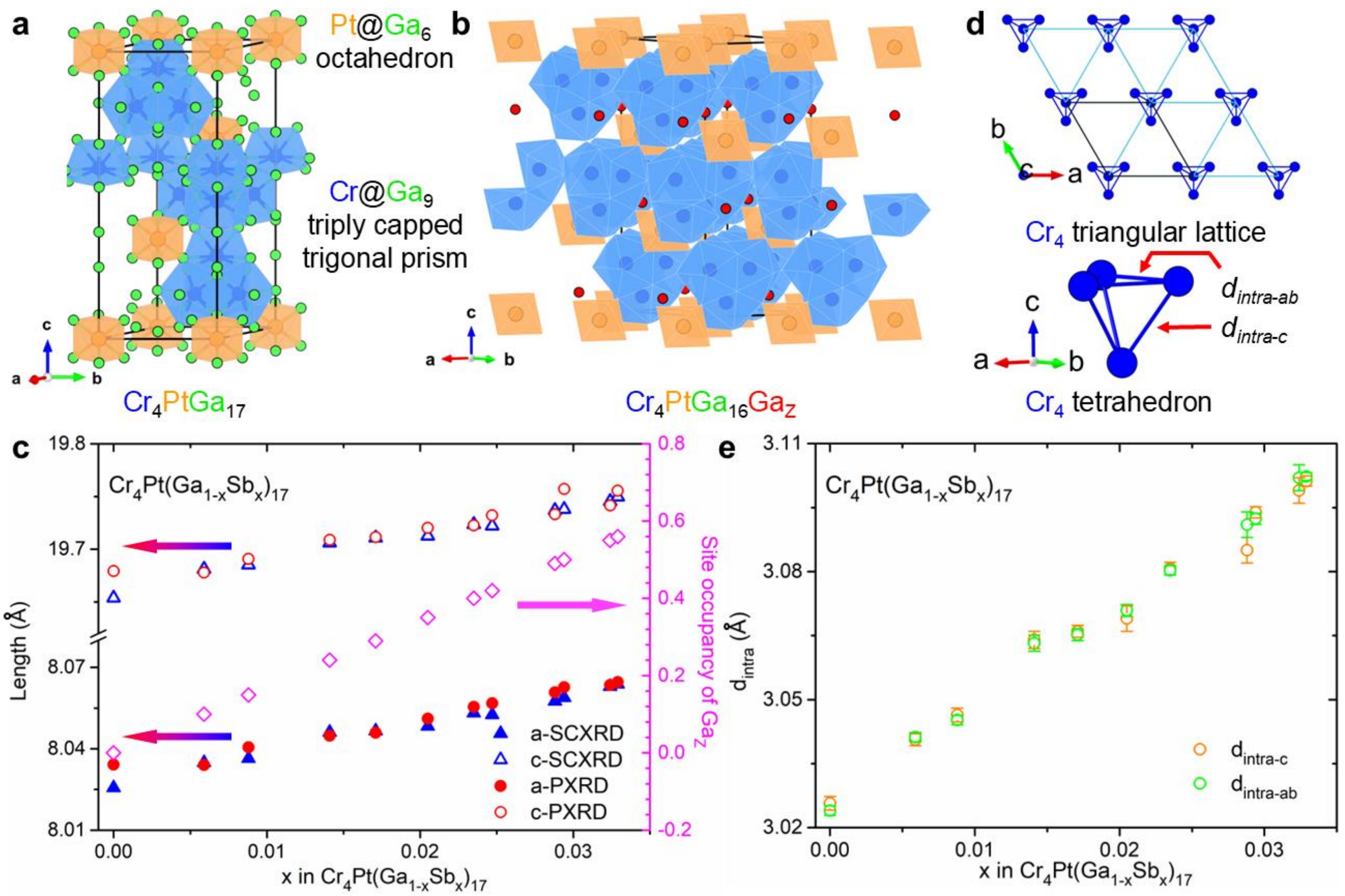

**Figure 1. a.** Crystal structure of $Cr_4PtGa_{17}$. Blue, orange and green circles represent Cr, Pt and Ga atoms. The coordination of Cr and Pt atoms with Ga atoms are shown in blue and orange polyhedral. **b.** Crystal structure of a 2×2×1 supercell of $Cr_4PtGa_{16}Ga_Z$ with $Ga_Z$ site emphasized in red. Other Ga sites are omitted. **c.** Changes of lattice parameters with Sb doping in $Cr_4Pt(Ga_{1-x}Sb_x)_{17}$ and the correspondence between x in $Cr_4Pt(Ga_{1-x}Sb_x)_{17}$ and site occupancy of $Ga_Z$ site. **d.** Triangular framework of $Cr_4$ tetrahedra and the orientation of each $Cr_4$ tetrahedron. **e.** Evolution of Cr-Cr intra-tetrahedral distances within *ab* plane and along *c* axis. All the data about undoped CPG are extracted from Ref. 27.

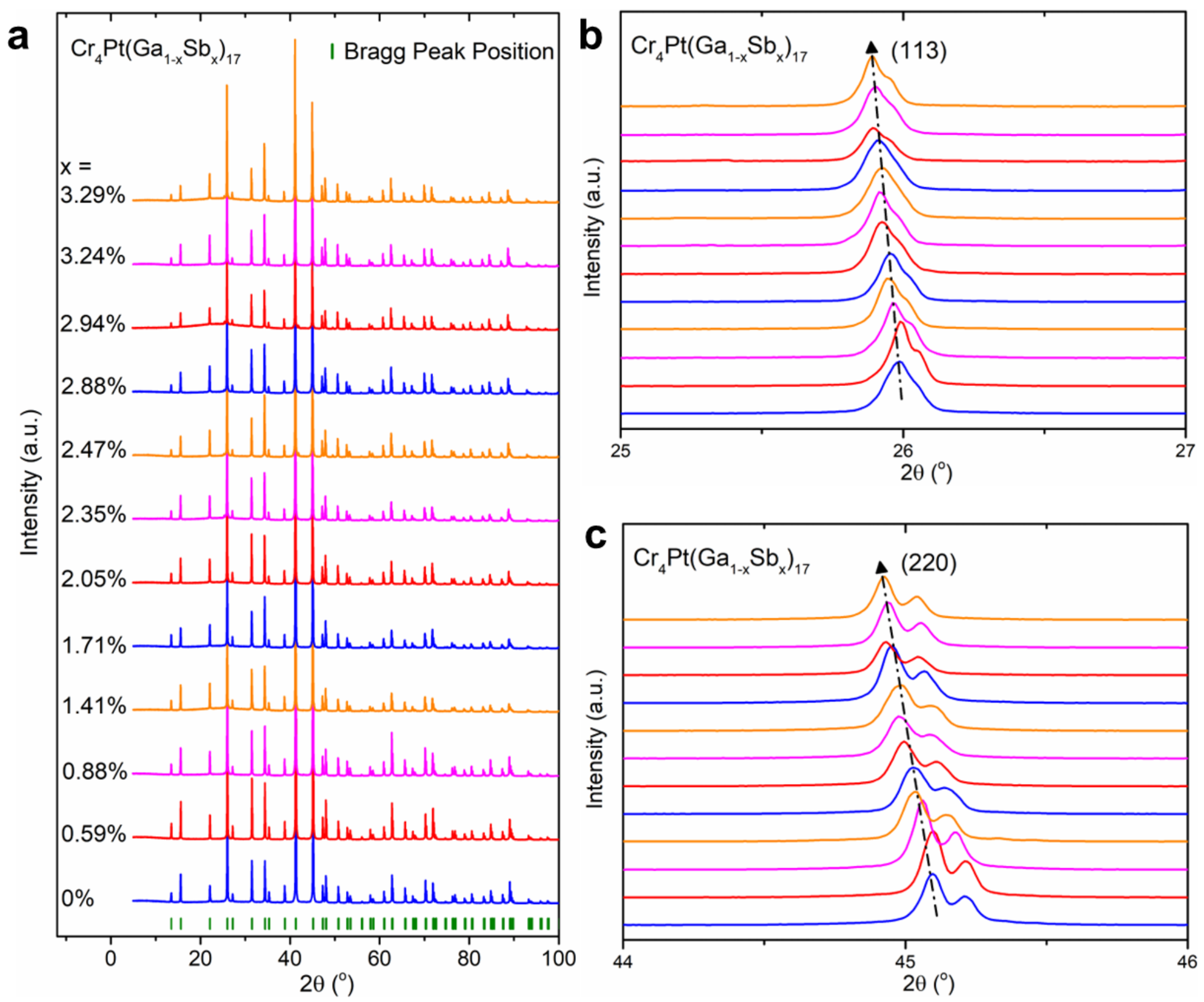

**Figure 2. a.** Powder X-ray diffraction patterns of all $Cr_4Pt(Ga_{1-x}Sb_x)_{17}$ samples. The green vertical ticks indicate the Bragg peak positions. **b. & c.** The peak position evolution of (113) and (220) planes.

The Cr atoms in CPG form a breathing pyrochlore lattice, which can be treated as a triangular lattice formed by $Cr_4$ tetrahedra, as shown in Figure 1d. Due to the rhombohedral symmetry, $Cr_4$ tetrahedra are tilted so that the Cr-Cr distance within the *ab* plane, i.e., $d_{intra-ab}$, is slightly different from the out-of-plane distance, i.e., $d_{intra-c}$. The impact of Sb doping on the intra-tetrahedral Cr-Cr distances is summarized in Figure 1e. Like the lattice parameters, a monotonical increase can be seen for both $d_{intra-ab}$ and $d_{intra-c}$ with Sb doping, as they both expand ~2.5 % from ~3.025 Å to ~3.102 Å.

Powder XRD was performed on crushed crystals of CPGS that were used for physical properties measurements. Rietveld refinement was completed on all. The obtained patterns are in good agreement with the reported CPG and the single crystal XRD results of CPGS in this report. No observable impurities can be seen in all patterns. The Bragg peaks shift monotonically towards lower 2θ, consistent with the lattice parameters' expansion with Sb doping. For instance, Figures 2b and 2c present the monotonical peak shift of two representative crystal planes, i.e., (113) and (220) planes.

**Evolution of magnetic properties with Sb doping:** All magnetic properties were measured on randomly aligned, multi-domain crystals of CPG and CPGS due to the unavailability of large single crystals for high-Sb-concentration samples to perform anisotropic measurements. Meanwhile, based on the reported nearly isotropic behavior of undoped CPG,[28] significant differences between magnetic behaviors on oriented single crystals and multi-domain crystals are not expected.

Figure 3a exhibits the evolution of temperature-dependent molar magnetic susceptibility measured under an external magnetic field of 0.1 T upon Sb doping on CPG in a field-cooling mode. The horizontal axis is shown in logarithmic type to emphasize low-temperature behavior. The magnetic susceptibility is defined as $\chi = {}^{M}/_{H}$ while the term attributed to core diamagnetism and temperature-independent paramagnetic contributions, $\chi_0$, has been subtracted. Ferromagnetic transition is observed at ~60 K for x = 0%, consistent with previous reports.[27,28] With increasing concentration of Sb, the ferromagnetic transition is gradually suppressed to lower temperature, as discussed in the later context. Meanwhile, the magnitude of magnetic susceptibility drops dramatically below 60 K from $\chi_{CPG}$ (1.8 K) = 4.34 emu/Oe/mol to $\chi_{CPGS-3.29\%}$ (1.8 K) = 0.03 emu/Oe/mol. To better understand how the magnetic behavior is affected by Sb doping, temperature-dependence of inverse magnetic susceptibility from 1.8 K to 100 K of CPGS are plotted in Figure 3b. Curie-Weiss (CW) law is applied to fit all the visible paramagnetic regions by using $\chi = \chi_0 + {}^{C}/_{(T - \theta_{CW})}$ where C is temperature-independent and is related to the effective moment ($\mu_{eff}$) via $\mu_{eff} = \sqrt{8C}$ and $\theta_{CW}$ is the CW temperature. The fitted $\theta_{CW}$ and $\mu_{eff}$ are summarized in Figure 3c. Monotonical decreasing of both $\theta_{CW}$ and $\mu_{eff}$ is observed while the $\theta_{CW}$ decreases from $\theta_{CW-CPG}$ = 61 (1) K to $\theta_{CW-CPGS-3.24\%}$ = 1.4 (2) K. When the Sb concentration increases to 3.29%, $\theta_{CW-CPGS-3.29\%}$ becomes negative, i.e., -1.8 (1) K, indicating a transition from weak ferromagnetic interaction in CPGS-3.24% to weak antiferromagnetic interaction in CPGS-3.29%. The $\mu_{eff}$ remains ~1.8-2.2 $\mu_B$/f.u. till CPGS-2.05% followed by a rapid decrease to ~0.68 (2) $\mu_B$/f.u. for CPGS-3.29%. A summary of how the CW-fitted parameters and the structural parameters are altered with varying dopant concentration of Sb in CPGS is also shown in Table 1.

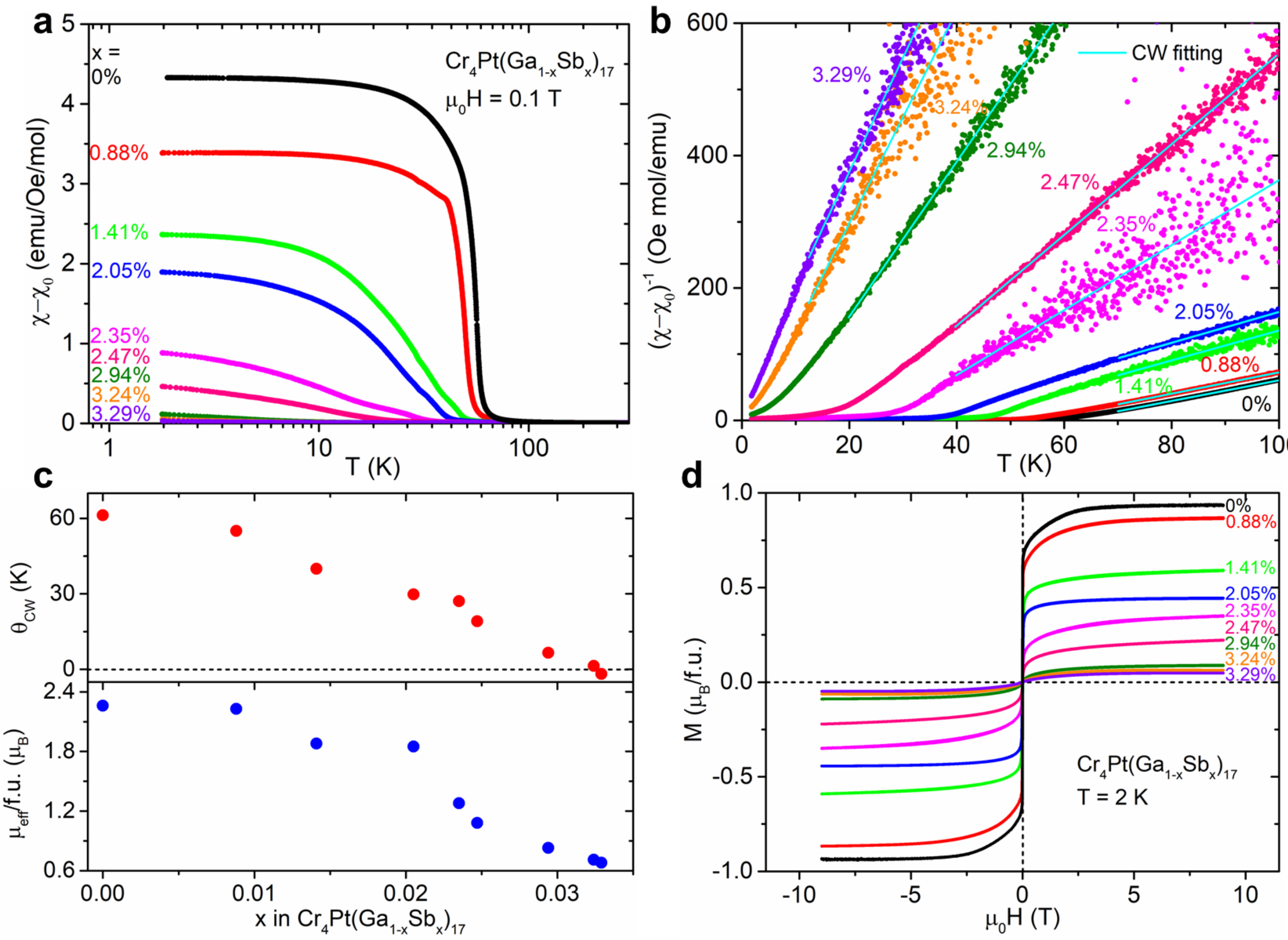

**Figure 3. a.** Temperature-dependence of modified magnetic susceptibility ($\chi$-$\chi_0$) for $Cr_4Pt(Ga_{1-x}Sb_x)_{17}$ under the external magnetic field of 0.1 T. The horizontal axis is shown in logarithmic scale. **b.** Temperature-dependent inverse modified magnetic susceptibility $(\chi$-$\chi_0)^{-1}$ for $Cr_4Pt(Ga_{1-x}Sb_x)_{17}$ under applied field of 0.1 T from 1.8 K to 100 K. The cyan lines indicate the Curie-Weiss (CW) fitting. **c.** The evolution of CW temperature ($\theta_{CW}$) and effective moment ($\mu_{eff}$) with increasing Sb doping. The dotted line stands for $\theta_{CW}$ = 0 K. **d.** Hysteresis loops of $Cr_4Pt(Ga_{1-x}Sb_x)_{17}$ from -9 T to 9 T at 2 K.

| x (%) | $\theta_{CW}$ (K) | $\mu_{eff}$/f.u. ($\mu_B$) | a (Å) | c (Å) | $d_{intra-ab}$ (Å) | $d_{intra-c}$ (Å) |
|---|---|---|---|---|---|---|
| **0** | 61.3 (2) | 2.26 (1) | 8.0256 (1) | 19.6539 (3) | 3.024 (1) | 3.026 (2) |
| **0.88** | 55.0 (2) | 2.23 (1) | 8.0364 (1) | 19.6852 (3) | 3.045 (1) | 3.046 (2) |
| **1.41** | 40.0 (7) | 1.88 (2) | 8.0461 (1) | 19.7062 (5) | 3.063 (2) | 3.064 (2) |
| **2.05** | 29.76 (1) | 1.85 (1) | 8.0483 (3) | 19.713 (1) | 3.071 (1) | 3.069 (3) |
| **2.35** | 26 (2) | 1.3 (1) | 8.0532 (1) | 19.7235 (5) | 3.080 (1) | 3.081 (2) |
| **2.47** | 19.13 (3) | 1.08 (2) | 8.0526 (1) | 19.7222 (4) | 3.081 (1) | 3.081 (1) |
| **2.94** | 6.6 (2) | 0.83 (1) | 8.0589 (1) | 19.7379 (4) | 3.092 (1) | 3.094 (1) |
| **3.24** | 1.4 (6) | 0.71 (2) | 8.0630 (4) | 19.745 (1) | 3.102 (3) | 3.099 (3) |
| **3.29** | -1.8 (4) | 0.68 (1) | 8.0638 (1) | 19.7498 (3) | 3.102 (1) | 3.101 (1) |

**Table 1.** Evolution of structural and magnetic parameters with dopant level of Sb in CPG.

Hysteresis loops of CPG and CPGS measured under 2 K are shown in Figure 3d. Consistent with Figure 3a, typical soft ferromagnetic behavior is observed for CPG and lightly doped CPGS. With Sb doping, the magnetic moment at 9 T ($\mu_{9T}$) is gradually suppressed from $\mu_{9T\text{-}CPG}$ ~0.94 $\mu_B$/f.u. to $\mu_{9T\text{-}CPGS\text{-}3.29\%}$ ~0.05 $\mu_B$/f.u.

**Suppression of metallicity in CPGS:** Considering the itinerant-like magnetism in CPG due to its unexpected small magnetic moment, similar to other itinerant magnets such as $La_5Co_2Ge_3$,[30] it is interesting to investigate how increasing the electron counts *via* Sb doping can affect the electrical transport properties. The measurements were carried out on crystals sizing from ~3×3×1 mm$^3$ to ~0.4×0.4×0.2 mm$^3$. Temperature-dependent resistivity of four representative samples and undoped CPG measured under no external magnetic field are shown in Figure 4a. With Sb doping, CPGS first becomes a "bad metal" at CPGS-1.41% due to the increasing resistivity throughout the whole temperature range compared to undoped CPG. The electrical transport behavior then exhibits semiconducting nature after the Sb concentration exceeds 2.05%, indicating the suppression of metallicity and a metal-semiconductor transition upon Sb doping. The resistivity of CPGS-3.24% is fitted using $\rho = \rho_0 e^{E/2k_BT}$, where $\rho_0$ is the pre-exponential term and $k_B$ is Boltzmann's constant, as shown in Figure 4b. The fitted bandgap of CPGS-3.24% is ~0.045 eV, which can be attributed to the charge carriers due to the electron-doping-induced impurity levels.

**Electronic structures of CPG and CPGS-5.89%:** According to the electrical resistivity of CPG and CPGS, it is expected that the Fermi energy ($E_F$) lies within a bandgap for CPGS where x > 2.05%, thus showing a semiconducting behavior. Figures 5a and 5b show the electronic band structure and density of states (DOS) of undoped CPG with consideration of both SOC on Pt atoms and ferromagnetism on Cr atoms, and CPGS-5.89%, for which the occupancy of Sb on the $Ga_Z$ site is 100% and SOC is taken into account for Pt and Sb. Although our experimental efforts have never reached the dopant level of 5.89%, this extreme case can directly show us how the $E_F$ and electronic structure evolve with Sb doping in CPG. It can be seen that for undoped CPG, the $E_F$ is in proximity to the top of the valence band of one spin channel and within a bandgap of the other spin channel, thus, exhibiting half-metallic nature.[27] In order to interpret the semiconducting behavior shown in Figure 4a, the band structure and DOS of CPGS-5.89% are shown in Figure 5b. No magnetic ordering is applied to the model while spin-orbit coupling (SOC) is employed only for Pt and Sb atoms. As shown, the $E_F$ lies on the bottom of a band, which corresponds to the band above bandgap in CPG. By comparing Figures 5a and 5b, it can be seen that the $E_F$ increases and results in metallic behavior when the dopant level is 5.89%, which is consistent with the fact that Sb possesses more electrons than Ga. Note that our experimental results revealed that the highest dopant level of Sb is 3.29 (5)%, corresponding to 56 (1)% occupancy on the $Ga_Z$ site. Therefore, we can speculate that when x > 2.05%, the $E_F$ increases to the top of the bottom band and the material exhibits electron-doped semiconducting behavior with increasing Sb concentration until x = 3.29%. As a result, the electrical resistivity shows a metal-semiconductor transition. Therefore, the semiconducting behvaior in CPGS can help explain the suppression of ferromagnetism in CPG, which is discussed later.

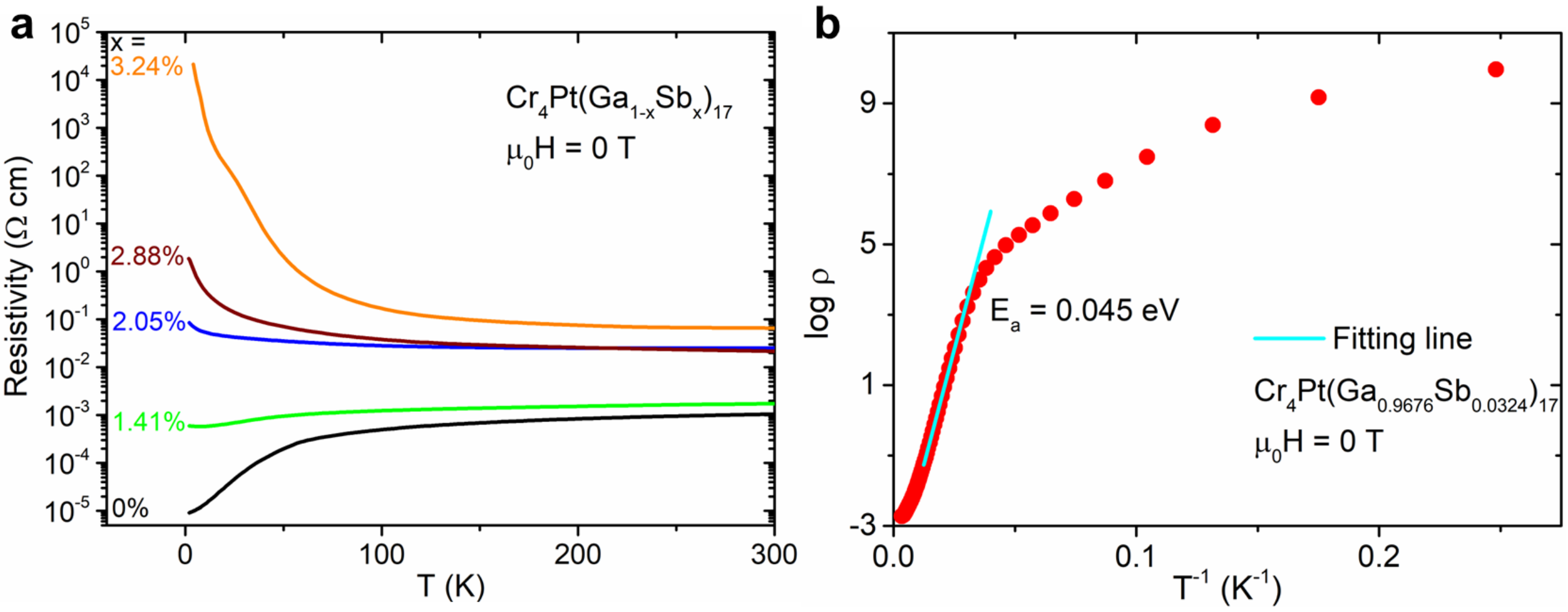

**Figure 4. a.** Temperature-dependent electrical resistivity of five representative samples of $Cr_4Pt(Ga_{1-x}Sb_x)_{17}$ (x = 0%, 1.41%, 2.05%, 2.88% and 3.24%) under zero applied magnetic field from 1.8 K to 300 K. The vertical axis is shown in logarithmic scale for better comparison. **b.** Logarithmic resistivity vs inverse temperature with fitting using Arrhenius equation.

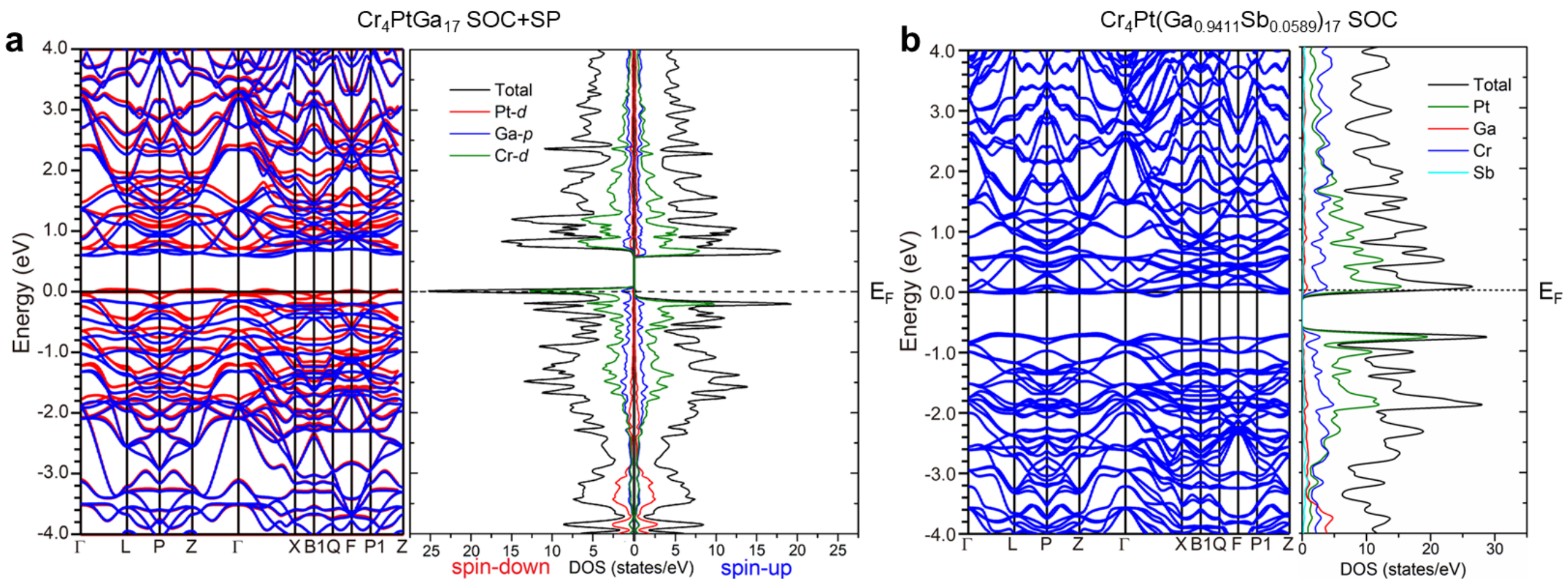

**Figure 5.** Electronic band structure and DOS of **a.** undoped ferromagnetic $Cr_4PtGa_{17}$ with consideration of SOC and spin polarization (SP) and **b.** $Cr_4Pt(Ga_{0.9411}Sb_{0.0589})_{17}$ with consideration of SOC only.

**Low-temperature heat capacity of CPGS:** To investigate if magnetic ordering exists below 2 K for high-Sb samples, zero-field temperature-dependence of heat capacity ($C_p$) measurements were conducted on small crystals of CPGS-2.94%, CPGS-3.24% and CPGS-3.29%, respectively. As shown in the inset of Figure 6a, no obvious anomaly corresponding to long-range magnetic ordering can be observed down to ~0.4 K for all three samples. The main panel of Figure 6a shows the $C_p/T$ vs T results. Slight upturn can be found for CPGS-2.94% and CPGS-3.24% while a significant upturn is seen for CPGS-3.29%, which can be attributed to Schottky anomaly. To confirm, $C_p/T$ vs $T^2$ curves are plotted in the main panel of Figure 6b for all three samples. The heat capacity of a material can usually be expressed as $C_p = \gamma T + \beta T^3$ where $\gamma$ and $\beta$ are electronic ($C_{el}$) and phononic ($C_{ph}$) contributions, respectively. For a semiconducting system like the three samples measured here, electronic contribution can be neglected. Therefore, the heat capacity can be expressed as $C_p = \beta T^3$, i.e., $C_p/T = \beta T^2$, which, however, does not lead to a good fitting between the linear relation of $C_p/T$ and $T^2$, and the observed data. Thus, phononic terms with higher orders are needed. Here, we conclude with $C_p/T = \beta_1 T^2 + \beta_2 T^4 + \beta_3 T^6 + \beta_4 T^8 + \beta_5 T^{10}$ where $\beta_1$-$\beta_5$ correspond to $C_{ph}$. The fitting parameters are summarized in Table S5 in the SI. The inset of Figure 6b illustrates the results after subtracting $C_{ph}$ from $C_p$. If the upturns for all three samples originate from magnonic contributions ($C_{mag}$), the magnetic entropy change ($\Delta S_{mag}$) can be obtained by integrating $C_{mag}/T$ vs T curves, as shown in Figure

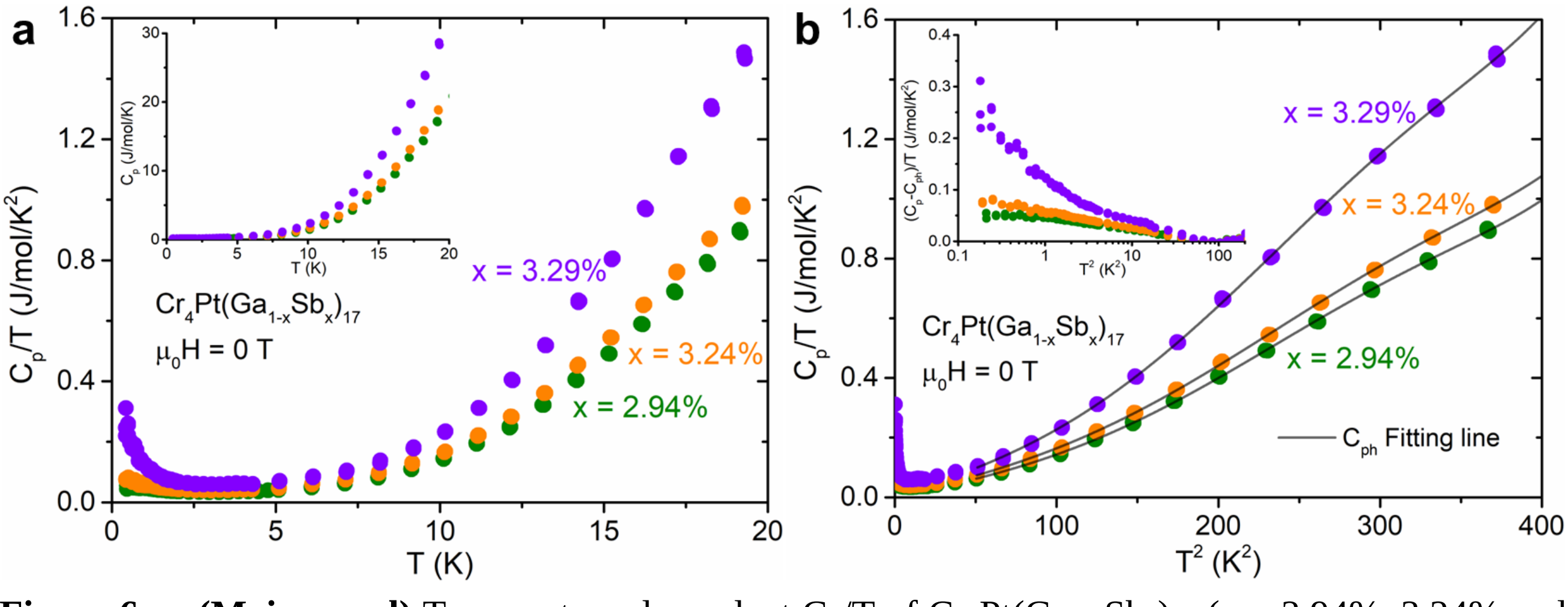

**Figure 6. a. (Main panel)** Temperature-dependent $C_p/T$ of $Cr_4Pt(Ga_{1-x}Sb_x)_{17}$ (x = 2.94%, 3.24% and 3.29%) under no applied magnetic field from 0.4 K to 20 K. **(Inset)** Temperature-dependent heat capacity ($C_p$) for the same samples from 0.4 K to 20 K. **b. (Main panel)** $C_p/T$ vs $T^2$ of $Cr_4Pt(Ga_{1-x}Sb_x)_{17}$ (x = 2.94%, 3.24% and 3.29%) under no applied magnetic field from 0.16 $K^2$ to 400 $K^2$. **(Inset)** Phonon-subtracted $C_p/T$, i.e., $(C_p - C_{ph})/T$, vs $T^2$ for all three samples. The horizontal axis is shown in logarithmic scale.

S1 in the SI. The proposed $\Delta S_{mag}$ for all three samples are smaller than 0.33 J/mol/K, which is only ~5.7% of Rln2 for a hypothetical $S$ = ½ system. Considering the lack of the λ-shape anomalies in the heat capacity measurements, they cannot be considered as long-range magnetic ordering within the measured temperature range.

**Discussion:** Based on the results shown above, one can tell that the magnetism in CPG more likely originates from itinerant electrons, i.e., itinerant magnetism, because the ferromagnetism in CPG is suppressed upon electron-doping. While the $E_F$ is at the edge of the bandgap for CPG, with Sb doping, the $E_F$ moves upwards into the bandgap. This makes the system enter a semiconducting region where there is no itinerant electron that can contribute to magnetism. One similar system is the single crystals of Mn-doped $SrRuO_3$ [37] where the parent compound, $SrRuO_3$, is a metallic itinerant ferromagnet. Upon Mn doping, the itinerant ferromagnetism is suppressed and insulating antiferromagnetism is induced at a critical concentration of Mn of ~39%. In this system, the Mn doping alters the Ru-Ru network and reduces the ferromagnetic coupling via the double exchange. With more Mn, superexchange mechanism prevailed, mediated by O ions, and the compounds become antiferromagnetic. Another example is Fe doped $SrRuO_3$ [38] where itinerant ferromagnetism is suppressed due to the large disorder on Ru site because of randomly distributed Fe ions. In our system of CPGS, we observed a slight antiferromagnetic coupling at the highest Sb concentration, which can be a potential onset of the insulting antiferromagnetic ordering. However, the solubility of Sb limits our investigation here and a different synthetic route, or another dopant will be needed to further study the magnetism beyond.

## *Conclusion*

Here, we systematically investigated the evolution of crystal structure, magnetism and electrical resistivity of $Cr_4Pt(Ga_{1-x}Sb_x)_{17}$ ($0 \leq x \leq 3.29$ (5)%). We have found that Sb doping can occur selectively on only one of the seven Ga sites in undoped $Cr_4PtGa_{17}$. The lattice parameters, Cr-Cr intra-tetrahedral distances and $Cr_4$-$Cr_4$ inter-tetrahedral distances increase monotonically with increasing Sb concentration. Moreover, the magnetic properties change dramatically with Sb doping. The ferromagnetism was found to be completely suppressed while the fitted Curie-Weiss temperatures and effective moments drop with more Sb doping. The absence of magnetism for samples with high Sb concentrations lasts till ~ 0.4 K, confirmed by temperature-dependent heat capacity. The ferromagnetism in $Cr_4PtGa_{17}$ is believed to be itinerant since the suppression of ferromagnetism is accompanied by the disappearance of metallicity, based on the electrical resistivity results. Sb-doped $Cr_4PtGa_{17}$ provides a deeper understanding of the magnetic behavior of this system and more studies are necessary to study its itinerant nature of ferromagnetism.

## *Acknowledgements*

C.W., G.A. and X.G. thank the support from the startup fund from the University of Pittsburgh and the Pitt Momentum Fund. Work at the University of Arizona is supported by the National Science Foundation under Award No. DMR-2338229.

## *Appendix A. Supplementary data*

Supplementary data to this article can be found online at xxxxxxx.

*Supporting Information*

# Evolution of Ferromagnetism and Electrical Resistivity in Sb-Doped $Cr_4PtGa_{17}$

*Chaoguo Wang,[a] Gina Angelo,[a] Jeremy G. Philbrick,[b] Tai Kong [b,c] and Xin Gui [a]**

[a] Department of Chemistry, University of Pittsburgh, Pittsburgh, PA, 15260, USA
[b] Department of Physics, University of Arizona, Tucson, AZ, 85721, USA
[c] Department of Chemistry and Biochemistry, University of Arizona, Tucson, AZ, 85721, USA

## Table of Contents

**Table S1.** The loading ratio of starting materials and observed composition for Sb in $Cr_4Pt(Ga_{1-x}Sb_x)_{17}$.

| **x** | **0.5%** | **1%** | **2%** | **3%** | **4%** | **5%** | **7%** | **10%** | **13%** | **16%** | **20%** |
|---|---|---|---|---|---|---|---|---|---|---|---|
| **Observed** | 0.59 (5)% | 0.88 (6)% | 1.41 (7)% | 1.71 (5)% | 2.05 (6)% | 2.35 (5)% | 2.47 (6)% | 2.88 (12)% | 2.94 (6)% | 3.24 (5)% | 3.29 (5)% |

**Table S2.** Single crystal structure refinement for $Cr_4PtGa_{16.90(1)}Sb_{0.10(1)}$, $Cr_4PtGa_{16.85(1)}Sb_{0.15(1)}$, $Cr_4PtGa_{16.74(1)}Sb_{0.26(1)}$, $Cr_4PtGa_{16.71(1)}Sb_{0.29(1)}$, $Cr_4PtGa_{16.65(1)}Sb_{0.35(1)}$, $Cr_4PtGa_{16.60(1)}Sb_{0.40(1)}$, $Cr_4PtGa_{16.58(1)}Sb_{0.42(1)}$, $Cr_4PtGa_{16.51(2)}Sb_{0.49(2)}$, $Cr_4PtGa_{16.50(1)}Sb_{0.50(1)}$, $Cr_4PtGa_{16.45(1)}Sb_{0.55(1)}$ and $Cr_4PtGa_{16.46(1)}Sb_{0.56(1)}$.

| Refined Formula | **$Cr_4PtGa_{16.90(1)}Sb_{0.10(1)}$** | **$Cr_4PtGa_{16.85(1)}Sb_{0.15(1)}$** |
|---|---|---|
| Temperature (K) | 293 (2) | 292 (2) |
| F.W. (g/mol) | 1593.71 | 1595.96 |
| Space group; Z | *R* 3m; 3 | *R* 3m; 3 |
| *a*(Å) | 8.03490 (7) | 8.03640 (7) |
| *c*(Å) | 19.6813 (3) | 19.6852 (3) |
| V (Å$^3$) | 1100.39 (3) | 1101.02 (3) |
| θ range (°) | 3.105-34.978 | 3.104-34.970 |
| No. reflections; $R_{int}$ | 13713; 0.0436 | 14350; 0.0422 |
| No. independent reflections | 1252 | 1260 |
| No. parameters | 52 | 52 |
| $R_1$: $\omega R_2$ ($I$>2δ($I$)) | 0.0148; 0.0309 | 0.0143; 0.0293 |
| Goodness of fit | 1.114 | 1.100 |
| Diffraction peak and hole (e$^-$/ Å$^3$) | 2.235; -1.735 | 0.996; -1.787 |
| Absolute structure parameter | 0.03 (1) | 0.02 (1) |

| Refined Formula | **$Cr_4PtGa_{16.74(1)}Sb_{0.26(1)}$** | **$Cr_4PtGa_{16.71(1)}Sb_{0.29(1)}$** | **$Cr_4PtGa_{16.65(1)}Sb_{0.35(1)}$** |
|---|---|---|---|
| Temperature (K) | 293 (2) | 293 (2) | 294 (2) |
| F.W. (g/mol) | 1600.99 | 1603.42 | 1606.37 |
| Space group; Z | *R* 3m; 3 | *R* 3m; 3 | *R* 3m; 3 |
| *a*(Å) | 8.0461 (1) | 8.0466 (2) | 8.0483 (3) |
| *c*(Å) | 19.706 (1) | 19.711 (1) | 19.713 (1) |
| V (Å$^3$) | 1104.85 (5) | 1105.25 (7) | 1105.82 (10) |
| θ range (°) | 3.101-34.926 | 3.100-34.919 | 3.100-34.913 |
| No. reflections; $R_{int}$ | 15784; 0.0602 | 14082; 0.0458 | 15369; 0.0670 |
| No. independent reflections | 1258 | 1259 | 1259 |
| No. parameters | 52 | 52 | 52 |
| $R_1$: $\omega R_2$ ($I$>2δ($I$)) | 0.0177; 0.0330 | 0.0144; 0.0289 | 0.0215; 0.0435 |
| Goodness of fit | 0.987 | 1.090 | 1.058 |
| Diffraction peak and hole (e$^-$/ Å$^3$) | 0.929; -0.780 | 1.035; -0.941 | 2.143; -2.092 |
| Absolute structure parameter | 0.02 (1) | 0.03 (1) | 0.02 (1) |

| Refined Formula | **$Cr_4PtGa_{16.60(1)}Sb_{0.40(1)}$** | **$Cr_4PtGa_{16.58(1)}Sb_{0.42(1)}$** | **$Cr_4PtGa_{16.51(2)}Sb_{0.49(2)}$** |
|---|---|---|---|
| Temperature (K) | 293 (2) | 293 (2) | 292 (2) |
| F.W. (g/mol) | 1609.14 | 1610.18 | 1613.65 |
| Space group; Z | *R* 3m; 3 | *R* 3m; 3 | *R* 3m; 3 |
| *a*(Å) | 8.0532 (1) | 8.0526 (1) | 8.0575 (1) |
| *c*(Å) | 19.7235 (5) | 19.7222 (4) | 19.7370 (6) |
| V (Å$^3$) | 1107.77 (5) | 1107.54 (4) | 1109.72 (5) |
| θ range (°) | 3.098-34.891 | 3.098-35.073 | 3.096-35.043 |
| No. reflections; $R_{int}$ | 14528; 0.0501 | 15171; 0.0483 | 13737; 0.0757 |
| No. independent reflections | 1261 | 1283 | 1266 |
| No. parameters | 52 | 52 | 52 |
| $R_1$: $\omega R_2$ ($I$>2δ($I$)) | 0.0150; 0.0274 | 0.0191; 0.0441 | 0.0243; 0.0499 |
| Goodness of fit | 0.976 | 1.164 | 1.067 |
| Diffraction peak and hole ($e^-$/ Å$^3$) | 0.972; -1.069 | 2.369; -2.891 | 1.257; -2.419 |
| Absolute structure parameter | 0.01 (1) | 0.03 (1) | 0.01 (1) |

| Refined Formula | **$Cr_4PtGa_{16.50(1)}Sb_{0.50(1)}$** | **$Cr_4PtGa_{16.45(1)}Sb_{0.55(1)}$** | **$Cr_4PtGa_{16.46(1)}Sb_{0.56(1)}$** |
|---|---|---|---|
| Temperature (K) | 292 (2) | 293 (2) | 295 (2) |
| F.W. (g/mol) | 1614.17 | 1617.12 | 1617.29 |
| Space group; Z | *R* 3m; 3 | *R* 3m; 3 | *R* 3m; 3 |
| *a*(Å) | 8.0589 (1) | 8.0630 (4) | 8.0638 (1) |
| *c*(Å) | 19.7379 (4) | 19.745 (1) | 19.7498 (3) |
| V (Å$^3$) | 1110.15 (3) | 1111.7 (1) | 1112.17 (3) |
| θ range (°) | 3.096-35.040 | 3.094-35.025 | 3.094-35.016 |
| No. reflections; $R_{int}$ | 13781; 0.0394 | 14110; 0.0662 | 13211; 0.0374 |
| No. independent reflections | 1262 | 1266 | 1277 |
| No. parameters | 52 | 52 | 52 |
| $R_1$: $\omega R_2$ ($I$>2δ($I$)) | 0.0129; 0.0248 | 0.0185; 0.0360 | 0.0128; 0.0251 |
| Goodness of fit | 0.967 | 0.974 | 0.901 |
| Diffraction peak and hole ($e^-$/ Å$^3$) | 0.874; -0.839 | 0.956; -1.372 | 0.935; -0.878 |
| Absolute structure parameter | 0.03 (1) | 0.98 (1) | 0.02 (1) |

**Table S3**. Atomic coordinates and equivalent isotropic displacement parameters for $Cr_4PtGa_{16.90(1)}Sb_{0.10(1)}$, $Cr_4PtGa_{16.85(1)}Sb_{0.15(1)}$, $Cr_4PtGa_{16.74(1)}Sb_{0.26(1)}$, $Cr_4PtGa_{16.71(1)}Sb_{0.29(1)}$, $Cr_4PtGa_{16.65(1)}Sb_{0.35(1)}$, $Cr_4PtGa_{16.60(1)}Sb_{0.40(1)}$, $Cr_4PtGa_{16.58(1)}Sb_{0.42(1)}$, $Cr_4PtGa_{16.51(2)}Sb_{0.49(2)}$, $Cr_4PtGa_{16.50(1)}Sb_{0.50(1)}$, $Cr_4PtGa_{16.45(1)}Sb_{0.55(1)}$ and $Cr_4PtGa_{16.46(1)}Sb_{0.56(1)}$ are defined as one-third of the trace of the orthogonalized $U_{ij}$ tensor (Å$^2$))

$Cr_4PtGa_{16.90(1)}Sb_{0.10(1)}$:

| **Atom** | **Wyck.** | **Occ.** | ***x*** | ***y*** | ***Z*** | $U_{eq}$ |
|---|---|---|---|---|---|---|
| Pt1 | 3*a* | 1 | 2/3 | 1/3 | 0.21554 (2) | 0.0066 (1) |
| Ga1 | 3*a* | 0.90 (1) | 0 | 0 | 0.13216 (5) | 0.0100 (3) |
| Sb1 | 3*a* | 0.10 (1) | 0 | 0 | 0.13216 (5) | 0.0100 (3) |
| Ga2 | 9*b* | 1 | 0.5236 (1) | 0.47643 (5) | 0.14397 (3) | 0.0119 (1) |
| Ga3 | 9*b* | 1 | 0.3806 (1) | 0.19028 (5) | 0.28701 (3) | 0.0118 (1) |
| Ga4 | 9*b* | 1 | 0.2112 (1) | 0.42239 (10) | 0.23776 (3) | 0.0124 (1) |
| Ga5 | 9*b* | 1 | 0.4555 (1) | 0.54455 (5) | 0.35988 (3) | 0.0125 (1) |
| Ga6 | 3*a* | 1 | 0 | 0 | 0.25504 (5) | 0.0106 (2) |
| Ga7 | 9*b* | 1 | 0.1638 (1) | 0.83619 (4) | 0.09122 (3) | 0.0106 (1) |
| Cr1 | 3*a* | 1 | 1/3 | 2/3 | 0.14338 (7) | 0.0065 (2) |
| Cr2 | 9*b* | 1 | 0.4595 (1) | 0.54050 (6) | 0.01725 (4) | 0.0066 (1) |

$Cr_4PtGa_{16.85(1)}Sb_{0.15(1)}$:

| **Atom** | **Wyck.** | **Occ.** | ***x*** | ***y*** | ***z*** | $U_{eq}$ |
|---|---|---|---|---|---|---|
| Pt1 | 3*a* | 1 | 0 | 0 | 0.79366 (2) | 0.0066 (1) |
| Ga1 | 3*a* | 0.85 (1) | 1/3 | 2/3 | 0.71030 (6) | 0.0098 (3) |
| Sb1 | 3*a* | 0.15 (1) | 1/3 | 2/3 | 0.71030 (6) | 0.0098 (3) |
| Ga2 | 3*a* | 1 | 1/3 | 2/3 | 0.83341 (6) | 0.0107 (2) |
| Ga3 | 9*b* | 1 | 0.85699 (5) | 0.14301 (5) | 0.72209 (4) | 0.0119 (1) |
| Ga4 | 9*b* | 1 | 0.14299 (5) | 0.85701 (5) | 0.86515 (3) | 0.0117 (1) |
| Ga5 | 9*b* | 1 | 0.54466 (5) | 0.45534 (5) | 0.81597 (3) | 0.0125 (1) |
| Ga6 | 9*b* | 1 | 0.49750 (4) | 0.99499 (9) | 0.66925 (4) | 0.0108 (1) |
| Ga7 | 9*b* | 1 | 0.78863 (5) | 0.21137 (5) | 0.93791 (4) | 0.0126 (1) |
| Cr1 | 3*a* | 1 | 2/3 | 1/3 | 0.72170 (8) | 0.0065 (3) |
| Cr2 | 9*b* | 1 | 0.45964 (6) | 0.54036 (6) | 0.92865 (5) | 0.0066 (1) |

$Cr_4PtGa_{16.74(1)}Sb_{0.26(1)}$:

| **Atom** | **Wyck.** | **Occ.** | ***x*** | ***y*** | ***z*** | $U_{eq}$ |
|---|---|---|---|---|---|---|
| Pt1 | 3*a* | 1 | 2/3 | 1/3 | 0.2537 (1) | 0.0056 (1) |
| Ga1 | 3*a* | 0.74 (1) | 2/3 | 1/3 | 0.0035 (1) | 0.0080 (3) |
| Sb1 | 3*a* | 0.26 (1) | 2/3 | 1/3 | 0.0035 (1) | 0.0080 (3) |
| Ga2 | 9*b* | 1 | 0.5237 (1) | 0.4763 (1) | 0.1822 (1) | 0.0110 (2) |
| Ga3 | 9*b* | 1 | 0.3808 (1) | 0.1904 (1) | 0.3251 (1) | 0.0110 (2) |
| Ga4 | 9*b* | 1 | 0.8785 (1) | 0.7570 (1) | 0.1095 (1) | 0.0117 (2) |
| Ga5 | 9*b* | 1 | 0.1677 (1) | 0.8322 (1) | 0.3783 (1) | 0.0100 (2) |
| Ga6 | 9*b* | 1 | 0.1216 (1) | 0.8784 (1) | 0.2311 (1) | 0.0118 (2) |
| Ga7 | 3*a* | 1 | 1/3 | 2/3 | 0.2128 (1) | 0.0102 (3) |
| Cr1 | 9*b* | 1 | 0.2538 (2) | 0.1269 (1) | 0.4520 (1) | 0.0058 (2) |
| Cr2 | 3*a* | 1 | 0 | 0 | 0.3250 (1) | 0.0059 (4) |

$Cr_4PtGa_{16.71(1)}Sb_{0.29(1)}$:

| **Atom** | **Wyck.** | **Occ.** | ***x*** | ***y*** | ***z*** | $U_{eq}$ |
|---|---|---|---|---|---|---|
| Pt1 | 3*a* | 1 | 1/3 | 2/3 | 0.32213 (2) | 0.0067 (1) |
| Ga1 | 3*a* | 0.71 (1) | 0 | 0 | 0.40540 (4) | 0.0094 (2) |
| Sb1 | 3*a* | 0.29 (1) | 0 | 0 | 0.40540 (4) | 0.0094 (2) |
| Ga2 | 3*a* | 1 | 0 | 0 | 0.28122 (5) | 0.0111 (2) |
| Ga3 | 9*b* | 1 | 0.19041 (5) | 0.3808 (1) | 0.25062 (3) | 0.0119 (1) |
| Ga4 | 9*b* | 1 | 0.47621 (5) | 0.9524 (1) | 0.39358 (3) | 0.0120 (1) |
| Ga5 | 9*b* | 1 | 0.83442 (4) | 1.1656 (1) | 0.44684 (3) | 0.0112 (1) |
| Ga6 | 9*b* | 1 | 0.21184 (5) | 0.7882 (1) | 0.51135 (3) | 0.0128 (1) |
| Ga7 | 9*b* | 1 | 0.78816 (5) | 1.5763 (1) | 0.29951 (3) | 0.0127 (1) |
| Cr1 | 9*b* | 1 | 0.87302 (6) | 0.2540 (1) | 0.18716 (4) | 0.0066 (1) |
| Cr2 | 3*a* | 1 | 2/3 | 4/3 | 0.39347 (7) | 0.0063 (2) |

$Cr_4PtGa_{16.65(1)}Sb_{0.35(1)}$:

| **Atom** | **Wyck.** | **Occ.** | ***x*** | ***y*** | ***z*** | $U_{eq}$ |
|---|---|---|---|---|---|---|
| Pt1 | 3*a* | 1 | 0 | 0 | 0.0883 (1) | 0.0050 (1) |
| Ga1 | 3*a* | 0.65 (1) | 2/3 | 1/3 | 0.1716 (1) | 0.0077 (4) |
| Sb1 | 3*a* | 0.35 (1) | 2/3 | 1/3 | 0.1716 (1) | 0.0077 (4) |
| Ga2 | 3*a* | 1 | 2/3 | 1/3 | 0.0470 (1) | 0.0095 (3) |
| Ga3 | 9*b* | 1 | 0.14282 (7) | 0.8572 (1) | 0.1597 (1) | 0.0104 (2) |
| Ga4 | 9*b* | 1 | 0.85708 (7) | 0.1429 (1) | 0.0169 (1) | 0.0102 (2) |
| Ga5 | 9*b* | 1 | 0.54466 (5) | 0.4553 (1) | 0.8160 (1) | 0.0125 (1) |
| Ga6 | 9*b* | 1 | 0.87872 (8) | 0.7574 (2) | 0.2777 (1) | 0.0112 (2) |
| Ga7 | 9*b* | 1 | 0.45464 (7) | 0.5454 (1) | 0.0657 (1) | 0.0113 (2) |
| Cr1 | 3*a* | 1 | 0.53948 (8) | 0.4605 (1) | 0.9534 (1) | 0.0048 (2) |
| Cr2 | 9*b* | 1 | 1/3 | 2/3 | 0.1597 (1) | 0.0051 (4) |

$Cr_4PtGa_{16.60(1)}Sb_{0.40(1)}$:

| **Atom** | **Wyck.** | **Occ.** | ***x*** | ***y*** | ***z*** | $U_{eq}$ |
|---|---|---|---|---|---|---|
| Pt1 | 3*a* | 1 | 0 | 0 | 0.94764 (2) | 0.0063 (1) |
| Ga1 | 3*a* | 0.60 (1) | 1/3 | 2/3 | 0.86424 (4) | 0.0084 (2) |
| Sb1 | 3*a* | 0.40 (1) | 1/3 | 2/3 | 0.86424 (4) | 0.0084 (2) |
| Ga2 | 3*a* | 1 | 1/3 | 2/3 | 0.98932 (6) | 0.0108 (2) |
| Ga3 | 9*b* | 1 | 0.85719 (5) | 0.1428 (1) | 0.87616 (3) | 0.0115 (1) |
| Ga4 | 9*b* | 1 | 0.14281 (5) | 0.8572 (1) | 0.01900 (3) | 0.0114 (1) |
| Ga5 | 9*b* | 1 | 0.12111 (5) | 0.2422 (1) | 0.75814 (3) | 0.0123 (1) |
| Ga6 | 9*b* | 1 | 0.50018 (5) | 0.0004 (1) | 0.82256 (3) | 0.0108 (1) |
| Ga7 | 9*b* | 1 | 0.54555 (5) | 0.0911 (1) | 0.97040 (3) | 0.0123 (1) |
| Cr1 | 3*a* | 1 | 2/3 | 1/3 | 0.87653 (7) | 0.0062 (3) |
| Cr2 | 9*b* | 1 | 0.46083 (6) | 0.5392 (1) | 0.08232 (5) | 0.0061 (1) |

$Cr_4PtGa_{16.58(1)}Sb_{0.42(1)}$:

| **Atom** | **Wyck.** | **Occ.** | ***x*** | ***y*** | ***z*** | $U_{eq}$ |
|---|---|---|---|---|---|---|
| Pt1 | 3*a* | 1 | 0 | 0 | 0.7701 (1) | 0.0057 (1) |
| Ga1 | 3*a* | 0.58 (1) | 2/3 | 1/3 | 0.8534 (1) | 0.0081 (3) |
| Sb1 | 3*a* | 0.42 (1) | 2/3 | 1/3 | 0.8534 (1) | 0.0081 (3) |
| Ga2 | 3*a* | 1 | 2/3 | 1/3 | 0.7282 (1) | 0.0101 (3) |
| Ga3 | 9*b* | 1 | 0.14281 (7) | 0.8572 (1) | 0.8415 (1) | 0.0109 (2) |
| Ga4 | 9*b* | 1 | 0.85717 (7) | 0.1428 (1) | 0.6986 (1) | 0.0110 (2) |
| Ga5 | 9*b* | 1 | 0.87895 (7) | 0.7579 (1) | 0.9595 (1) | 0.0117 (2) |
| Ga6 | 9*b* | 1 | 0.45432 (7) | 0.9086 (1) | 0.7472 (1) | 0.0118 (2) |
| Ga7 | 9*b* | 1 | 0.49974 (6) | 0.9995 (1) | 0.8951 (1) | 0.0102 (2) |
| Cr1 | 3*a* | 1 | 1/3 | 2/3 | 0.8410 (1) | 0.0056 (3) |
| Cr2 | 9*b* | 1 | 0.20578 (8) | 0.7942 (1) | 0.9686 (1) | 0.0055 (2) |

$Cr_4PtGa_{16.51(2)}Sb_{0.49(2)}$:

| **Atom** | **Wyck.** | **Occ.** | ***x*** | ***y*** | ***z*** | $U_{eq}$ |
|---|---|---|---|---|---|---|
| Pt1 | 3*a* | 1 | 0 | 0 | 0.0856 (1) | 0.0063 (1) |
| Ga1 | 3*a* | 0.51 (2) | 1/3 | 2/3 | 0.0022 (1) | 0.0081 (4) |
| Sb1 | 3*a* | 0.49 (2) | 1/3 | 2/3 | 0.0022 (1) | 0.0081 (4) |
| Ga2 | 9*b* | 1 | 0.1427 (1) | 0.8573 (1) | 0.1569 (1) | 0.0116 (3) |
| Ga3 | 9*b* | 1 | 0.8572 (1) | 0.1428 (1) | 0.0143 (1) | 0.0113 (2) |
| Ga4 | 3*a* | 1 | 1/3 | 2/3 | 0.1280 (1) | 0.0107 (4) |
| Ga5 | 9*b* | 1 | 0.7582 (2) | 0.8791 (1) | 0.8960 (1) | 0.0123 (2) |
| Ga6 | 9*b* | 1 | 0.1677 (1) | 0.8323 (1) | 0.2937 (1) | 0.0107 (2) |
| Ga7 | 9*b* | 1 | 0.5458 (1) | 0.0917 (2) | 0.1085 (1) | 0.0122 (2) |
| Cr1 | 9*b* | 1 | 0.4612 (1) | 0.9224 (2) | 0.2206 (1) | 0.0060 (3) |
| Cr2 | 3*a* | 1 | 2/3 | 1/3 | 0.0147 (2) | 0.0067 (5) |

$Cr_4PtGa_{16.50(1)}Sb_{0.50(1)}$:

| Atom | Wyck. | Occ. | *x* | *y* | *Z* | $U_{eq}$ |
|---|---|---|---|---|---|---|
| Pt1 | 3*a* | 1 | 0 | 0 | 0.02769 (2) | 0.0060 (1) |
| Ga1 | 3*a* | 0.50 (1) | 0 | 0 | 0.77775 (4) | 0.0077 (2) |
| Sb1 | 3*a* | 0.50 (1) | 0 | 0 | 0.77775 (4) | 0.0077 (2) |
| Ga2 | 9*b* | 1 | 0.8572 (1) | 0.14277 (4) | 0.95641 (3) | 0.0111 (1) |
| Ga3 | 9*b* | 1 | 0.2126 (1) | 0.42526 (9) | 0.88406 (3) | 0.0119 (1) |
| Ga4 | 9*b* | 1 | 0.8322 (1) | 0.16778 (4) | 0.81970 (3) | 0.0103 (1) |
| Ga5 | 9*b* | 1 | 0.1428 (1) | 0.85724 (4) | 0.09920 (3) | 0.0112 (1) |
| Ga6 | 9*b* | 1 | 0.7874 (1) | 0.21259 (4) | 0.67146 (3) | 0.0121 (1) |
| Ga7 | 3*a* | 1 | 2/3 | 1/3 | 0.98528 (5) | 0.0102 (2) |
| Cr1 | 3*a* | 1 | 2/3 | 1/3 | 0.76513 (6) | 0.0056 (2) |
| Cr2 | 9*b* | 1 | 0.2250 (1) | 0.46124 (5) | 0.89315 (4) | 0.0057 (1) |

$Cr_4PtGa_{16.45(1)}Sb_{0.55(1)}$:

| Atom | Wyck. | Occ. | *x* | *y* | *z* | $U_{eq}$ |
|---|---|---|---|---|---|---|
| Pt1 | 3*a* | 1 | 0 | 0 | 0.2886 (1) | 0.0062 (1) |
| Ga1 | 3*a* | 0.45 (1) | 1/3 | 2/3 | 0.2053 (1) | 0.0082 (3) |
| Sb1 | 3*a* | 0.55 (1) | 1/3 | 2/3 | 0.2053 (1) | 0.0082 (3) |
| Ga2 | 9*b* | 1 | 0.14266 (7) | 0.8573 (1) | 0.3600 (1) | 0.0117 (2) |
| Ga3 | 9*b* | 1 | 0.85726 (7) | 0.1427 (1) | 0.2173 (1) | 0.0116 (2) |
| Ga4 | 9*b* | 1 | 0.50169 (6) | 0.4983 (1) | 0.1632 (1) | 0.0107 (2) |
| Ga5 | 9*b* | 1 | 0.12063 (7) | 0.2413 (1) | 0.0989 (1) | 0.0124 (2) |
| Ga6 | 9*b* | 1 | 0.54605 (7) | 0.4540 (1) | 0.3117 (1) | 0.0123 (2) |
| Ga7 | 3*a* | 1 | 1/3 | 2/3 | 0.3315 (1) | 0.0105 (3) |
| Cr1 | 9*b* | 1 | 0.46158 (8) | 0.9231 (2) | 0.4233 (1) | 0.0064 (2) |
| Cr2 | 3*a* | 1 | 2/3 | 1/3 | 0.2181 (1) | 0.0061 (4) |

$Cr_4PtGa_{16.46(1)}Sb_{0.56(1)}$:

| Atom | Wyck. | Occ. | *x* | *y* | *z* | $U_{eq}$ |
|---|---|---|---|---|---|---|
| Pt1 | 3*a* | 1 | 0 | 0 | 0.75005 (2) | 0.0062 (1) |
| Ga1 | 3*a* | 0.44 (1) | 0 | 0 | 0.00000 (3) | 0.0080 (2) |
| Sb1 | 3*a* | 0.56 (1) | 0 | 0 | 0.00000 (3) | 0.0080 (2) |
| Ga2 | 9*b* | 1 | 0.80938 (4) | 0.19062 (4) | 0.15474 (3) | 0.0115 (1) |
| Ga3 | 9*b* | 1 | 0.52393 (4) | 0.47607 (4) | 0.01197 (3) | 0.0114 (1) |
| Ga4 | 9*b* | 1 | 0.45389 (4) | 0.54611 (4) | 0.22697 (3) | 0.0121 (1) |
| Ga5 | 9*b* | 1 | 0.21275 (4) | 0.78725 (4) | 0.10639 (3) | 0.0123 (1) |
| Ga6 | 9*b* | 1 | 0.83509 (4) | 0.16491 (4) | 0.29127 (3) | 0.0106 (1) |
| Ga7 | 3*a* | 1 | 0 | 0 | 0.12631 (5) | 0.0105 (2) |
| Cr1 | 9*b* | 1 | 0.12824 (5) | 0.87176 (5) | 0.21793 (4) | 0.0061 (1) |
| Cr2 | 3*a* | 1 | 0 | 0 | 0.34612 (6) | 0.0061 (2) |

**Table S4.** Anisotropic thermal displacement parameters for $Cr_4PtGa_{16.90(1)}Sb_{0.10(1)}$, $Cr_4PtGa_{16.85(1)}Sb_{0.15(1)}$, $Cr_4PtGa_{16.74(1)}Sb_{0.26(1)}$, $Cr_4PtGa_{16.71(1)}Sb_{0.29(1)}$, $Cr_4PtGa_{16.65(1)}Sb_{0.35(1)}$, $Cr_4PtGa_{16.60(1)}Sb_{0.40(1)}$, $Cr_4PtGa_{16.58(1)}Sb_{0.42(1)}$, $Cr_4PtGa_{16.51(2)}Sb_{0.49(2)}$, $Cr_4PtGa_{16.50(1)}Sb_{0.50(1)}$, $Cr_4PtGa_{16.45(1)}Sb_{0.55(1)}$ and $Cr_4PtGa_{16.46(1)}Sb_{0.56(1)}$.

$Cr_4PtGa_{16.90(1)}Sb_{0.10(1)}$:

| Atom | U11 | U22 | U33 | U12 | U13 | U23 |
|---|---|---|---|---|---|---|
| Pt1 | 0.0072 (1) | 0.0072 (1) | 0.0055 (1) | 0.0036 (1) | 0 | 0 |
| Ga1 | 0.0106 (3) | 0.0106 (3) | 0.0088 (4) | 0.0053 (2) | 0 | 0 |
| Sb1 | 0.0106 (3) | 0.0106 (3) | 0.0088 (4) | 0.0053 (2) | 0 | 0 |
| Ga2 | 0.0141 (2) | 0.0141 (2) | 0.0105 (3) | 0.0094 (2) | -0.0016(1) | 0.0016 (1) |
| Ga3 | 0.0102 (3) | 0.0131 (2) | 0.0111 (3) | 0.0051 (1) | 0.0039 (2) | 0.0020 (1) |
| Ga4 | 0.0148 (2) | 0.0132 (3) | 0.0087 (2) | 0.0066 (1) | 0.0019 (1) | 0.0037 (2) |
| Ga5 | 0.0098 (2) | 0.0098 (2) | 0.0140 (3) | 0.0020 (2) | 0.0001 (1) | -0.0001(1) |
| Ga6 | 0.0117 (3) | 0.0117 (3) | 0.0084 (4) | 0.0059 (1) | 0 | 0 |
| Ga7 | 0.0112 (2) | 0.0112 (2) | 0.0099 (2) | 0.0061 (2) | 0.0002 (1) | -0.0002(1) |
| Cr1 | 0.0069 (3) | 0.0069 (3) | 0.0056 (6) | 0.0035 (2) | 0 | 0 |
| Cr2 | 0.0071 (2) | 0.0071 (2) | 0.0053 (3) | 0.0032 (3) | -0.0001(1) | 0.0001 (1) |

$Cr_4PtGa_{16.85(1)}Sb_{0.15(1)}$:

| Atom | U11 | U22 | U33 | U12 | U13 | U23 |
|---|---|---|---|---|---|---|
| Pt1 | 0.0065 (1) | 0.0065 (1) | 0.0067 (1) | 0.0033 (1) | 0 | 0 |
| Ga1 | 0.0099 (3) | 0.0099 (3) | 0.0095 (5) | 0.0050 (2) | 0 | 0 |
| Sb1 | 0.0099 (3) | 0.0099 (3) | 0.0095 (5) | 0.0050 (2) | 0 | 0 |
| Ga2 | 0.0109 (3) | 0.0109 (3) | 0.0102 (5) | 0.0055 (1) | 0 | 0 |
| Ga3 | 0.0136 (2) | 0.0136 (2) | 0.0114 (3) | 0.0089 (3) | -0.0016(1) | 0.0016 (1) |
| Ga4 | 0.0125 (2) | 0.0125 (2) | 0.0122 (3) | 0.0078(2) | -0.0020(1) | 0.0020 (1) |
| Ga5 | 0.0145 (2) | 0.0145 (2) | 0.0099 (3) | 0.0083(2) | 0.0019 (1) | -0.0019(1) |
| Ga6 | 0.0109 (2) | 0.0103 (3) | 0.0109 (3) | 0.0052 (1) | 0.0001 (1) | 0.0002 (2) |
| Ga7 | 0.0092 (2) | 0.0092 (2) | 0.0152 (3) | 0.0016 (2) | 0.0001 (1) | -0.0001(1) |
| Cr1 | 0.0060 (4) | 0.0060 (4) | 0.0073 (7) | 0.0030 (2) | 0 | 0 |
| Cr2 | 0.0065 (2) | 0.0065 (2) | 0.0063 (3) | 0.0029 (3) | -0.0002(2) | 0.0002 (2) |

$Cr_4PtGa_{16.74(1)}Sb_{0.26(1)}$:

| Atom | U11 | U22 | U33 | U12 | U13 | U23 |
|---|---|---|---|---|---|---|
| Pt1 | 0.0057 (1) | 0.0057 (1) | 0.0055 (2) | 0.0028 (1) | 0 | 0 |
| Ga1 | 0.0079 (4) | 0.0079 (4) | 0.0083 (6) | 0.0039 (2) | 0 | 0 |
| Sb1 | 0.0079 (4) | 0.0079 (4) | 0.0083 (6) | 0.0039 (2) | 0 | 0 |
| Ga2 | 0.0116 (3) | 0.0116 (3) | 0.0117 (4) | 0.0018 (2) | -0.0018(2) | 0.0018 (2) |
| Ga3 | 0.0083 (4) | 0.0128 (3) | 0.0102 (4) | 0.0042 (2) | 0.0035 (3) | 0.0018 (2) |
| Ga4 | 0.0083 (3) | 0.0145 (4) | 0.0145 (4) | 0.0072 (2) | 0.0003 (2) | 0.0005 (3) |
| Ga5 | 0.0100 (3) | 0.0100 (3) | 0.0096 (4) | 0.0047 (3) | -0.0001(2) | 0.0001 (2) |
| Ga6 | 0.0137 (3) | 0.0137 (3) | 0.0090 (4) | 0.0075 (3) | 0.0021 (1) | -0.0021(1) |
| Ga7 | 0.0097 (4) | 0.0097 (4) | 0.0111 (7) | 0.0048 (2) | 0 | 0 |
| Cr1 | 0.0066 (5) | 0.0059 (3) | 0.0052 (5) | 0.0033 (2) | 0.0000 (4) | 0.0000 (2) |
| Cr2 | 0.0059 (5) | 0.0059 (5) | 0.0060 (9) | 0.0029 (3) | 0 | 0 |

$Cr_4PtGa_{16.71(1)}Sb_{0.29(1)}$:

| Atom | U11 | U22 | U33 | U12 | U13 | U23 |
|---|---|---|---|---|---|---|
| Pt1 | 0.0071 (1) | 0.0071 (1) | 0.0059 (1) | 0.0036 (1) | 0 | 0 |
| Ga1 | 0.0100 (3) | 0.0100 (3) | 0.0084 (4) | 0.0050 (2) | 0 | 0 |
| Sb1 | 0.0100 (3) | 0.0100 (3) | 0.0084 (4) | 0.0050 (2) | 0 | 0 |
| Ga2 | 0.0111 (3) | 0.0111 (3) | 0.0111 (5) | 0.0055 (1) | 0 | 0 |
| Ga3 | 0.0131 (2) | 0.0104 (3) | 0.0113 (3) | 0.0052 (1) | -0.0018(1) | -0.0036(2) |
| Ga4 | 0.0141 (2) | 0.0096 (3) | 0.0107 (3) | 0.0048 (1) | -0.0017(1) | -0.0033(2) |
| Ga5 | 0.0116 (2) | 0.0116 (2) | 0.0099 (2) | 0.0055 (2) | -0.0001(1) | 0.0001 (1) |
| Ga6 | 0.0098 (2) | 0.0098 (2) | 0.0144 (3) | 0.0017 (2) | 0.0002 (1) | -0.0002(1) |
| Ga7 | 0.0151 (2) | 0.0132 (3) | 0.0092 (3) | 0.0066 (1) | 0.0020 (1) | 0.0040 (2) |
| Cr1 | 0.0069 (2) | 0.0079 (3) | 0.0053 (3) | 0.0040 (2) | -0.0002(1) | -0.0003(3) |
| Cr2 | 0.0064 (4) | 0.0064 (4) | 0.0059 (6) | 0.0032 (2) | 0 | 0 |

$Cr_4PtGa_{16.65(1)}Sb_{0.35(1)}$:

| Atom | U11 | U22 | U33 | U12 | U13 | U23 |
|---|---|---|---|---|---|---|
| Pt1 | 0.0049 (1) | 0.0049 (1) | 0.0053 (2) | 0.0024 (1) | 0 | 0 |
| Ga1 | 0.0076 (4) | 0.0076 (4) | 0.0080 (6) | 0.0038 (2) | 0 | 0 |
| Sb1 | 0.0076 (4) | 0.0076 (4) | 0.0080 (6) | 0.0038 (2) | 0 | 0 |
| Ga2 | 0.0084 (4) | 0.0084 (4) | 0.0119 (8) | 0.0042 (2) | 0 | 0 |
| Ga3 | 0.0120 (3) | 0.0120 (3) | 0.0100 (4) | 0.0016 (2) | -0.0016(2) | 0.0016 (2) |
| Ga4 | 0.0110 (3) | 0.0110 (3) | 0.0108 (4) | 0.0071 (4) | -0.0020(2) | 0.0020 (2) |
| Ga5 | 0.0091 (3) | 0.0091 (3) | 0.0095 (4) | 0.0001 (2) | -0.0001(2) | 0.0001 (2) |
| Ga6 | 0.0076 (3) | 0.0140 (4) | 0.0141 (5) | 0.0070 (2) | 0.0002 (2) | 0.0003 (3) |
| Ga7 | 0.0128 (3) | 0.0128 (3) | 0.0088 (4) | 0.0069 (4) | 0.0021 (2) | -0.0021(2) |
| Cr1 | 0.0042 (4) | 0.0042 (4) | 0.0053 (5) | 0.0016 (4) | 0.0000 (2) | 0.0000 (2) |
| Cr2 | 0.0048 (5) | 0.0048 (5) | 0.0057 (10) | 0.0024 (3) | 0 | 0 |

$Cr_4PtGa_{16.60(1)}Sb_{0.40(1)}$:

| **Atom** | **U11** | **U22** | **U33** | **U12** | **U13** | **U23** |
|---|---|---|---|---|---|---|
| Pt1 | 0.0065 (1) | 0.0065 (1) | 0.0060 (1) | 0.0033 (1) | 0 | 0 |
| Ga1 | 0.0086 (3) | 0.0086 (3) | 0.0079 (4) | 0.0043 (1) | 0 | 0 |
| Sb1 | 0.0086 (3) | 0.0086 (3) | 0.0079 (4) | 0.0043 (1) | 0 | 0 |
| Ga2 | 0.0104 (3) | 0.0104 (3) | 0.0115 (5) | 0.0052 (1) | 0 | 0 |
| Ga3 | 0.0132 (2) | 0.0132 (2) | 0.0110 (3) | 0.0088 (3) | -0.0016(1) | 0.0016 (1) |
| Ga4 | 0.0127 (2) | 0.0127 (2) | 0.0111 (3) | 0.0079 (2) | -0.0018(1) | 0.0018 (1) |
| Ga5 | 0.0093 (2) | 0.0152 (3) | 0.0144 (3) | 0.0076 (2) | 0.0003 (1) | 0.0005 (2) |
| Ga6 | 0.0107 (2) | 0.0122 (3) | 0.0099 (3) | 0.0061 (1) | -0.0004(1) | -0.0007(2) |
| Ga7 | 0.0144 (2) | 0.0128 (3) | 0.0091 (3) | 0.0064 (1) | 0.0019 (1) | 0.0039 (2) |
| Cr1 | 0.0059 (4) | 0.0059 (4) | 0.0067 (6) | 0.0030 (2) | 0 | 0 |
| Cr2 | 0.0064 (2) | 0.0064 (2) | 0.0056 (3) | 0.0031 (3) | -0.0002(2) | 0.0002 (2) |

$Cr_4PtGa_{16.58(1)}Sb_{0.42(1)}$:

| **Atom** | **U11** | **U22** | **U33** | **U12** | **U13** | **U23** |
|---|---|---|---|---|---|---|
| Pt1 | 0.0059 (1) | 0.0059 (1) | 0.0053 (2) | 0.0029 (1) | 0 | 0 |
| Ga1 | 0.0084 (4) | 0.0084 (4) | 0.0075 (5) | 0.0042 (2) | 0 | 0 |
| Sb1 | 0.0084 (4) | 0.0084 (4) | 0.0075 (5) | 0.0042 (2) | 0 | 0 |
| Ga2 | 0.0096 (4) | 0.0096 (4) | 0.0110 (7) | 0.0048 (2) | 0 | 0 |
| Ga3 | 0.0126 (3) | 0.0126 (3) | 0.0102 (4) | 0.0085 (4) | -0.0018(1) | 0.0018 (1) |
| Ga4 | 0.0121 (3) | 0.0121 (3) | 0.0111 (4) | 0.0077 (3) | -0.0019(2) | 0.0019 (2) |
| Ga5 | 0.0088 (3) | 0.0147 (4) | 0.0137 (4) | 0.0074 (2) | 0.0002 (2) | 0.0004 (3) |
| Ga6 | 0.0137 (3) | 0.0125 (4) | 0.0086 (4) | 0.0062 (2) | 0.0020 (1) | 0.0039 (3) |
| Ga7 | 0.0102 (3) | 0.0118 (4) | 0.0091 (3) | 0.0059 (2) | -0.0003(2) | -0.0006(3) |
| Cr1 | 0.0056 (5) | 0.0056 (5) | 0.0055 (8) | 0.0028 (3) | 0 | 0 |
| Cr2 | 0.0056 (3) | 0.0056 (3) | 0.0049 (4) | 0.0026 (4) | 0.0000 (2) | -0.0000(2) |

$Cr_4PtGa_{16.51(2)}Sb_{0.49(2)}$:

| **Atom** | **U11** | **U22** | **U33** | **U12** | **U13** | **U23** |
|---|---|---|---|---|---|---|
| Pt1 | 0.0058 (2) | 0.0058 (2) | 0.0071 (3) | 0.0029 (1) | 0 | 0 |
| Ga1 | 0.0076 (5) | 0.0076 (5) | 0.0092 (7) | 0.0038 (2) | 0 | 0 |
| Sb1 | 0.0076 (5) | 0.0076 (5) | 0.0092 (7) | 0.0038 (2) | 0 | 0 |
| Ga2 | 0.0121 (4) | 0.0121 (4) | 0.0130 (6) | 0.0078 (5) | -0.0018(2) | 0.0018 (2) |
| Ga3 | 0.0126 (4) | 0.0126 (4) | 0.0113 (6) | 0.0081 (5) | -0.0017(2) | 0.0017 (2) |
| Ga4 | 0.0093 (5) | 0.0093 (5) | 0.0134 (11) | 0.0046 (3) | 0 | 0 |
| Ga5 | 0.0152 (6) | 0.0088 (4) | 0.0150 (6) | 0.0076 (3) | -0.0001(5) | -0.0001(2) |
| Ga6 | 0.0099 (4) | 0.0099 (4) | 0.0111 (5) | 0.0041 (4) | -0.0004(2) | 0.0004 (2) |
| Ga7 | 0.0139 (4) | 0.0124 (5) | 0.0099 (6) | 0.0062 (3) | 0.0020 (2) | 0.0040 (4) |
| Cr1 | 0.0058 (5) | 0.0065 (7) | 0.0060 (7) | 0.0033 (3) | -0.0002(3) | -0.0004(6) |
| Cr2 | 0.0056 (7) | 0.0056 (7) | 0.009 (1) | 0.0028 (4) | 0 | 0 |

$Cr_4PtGa_{16.50(1)}Sb_{0.50(1)}$:

| Atom | U11 | U22 | U33 | U12 | U13 | U23 |
|---|---|---|---|---|---|---|
| Pt1 | 0.0061 (1) | 0.0061 (1) | 0.0056 (1) | 0.0031 (1) | 0 | 0 |
| Ga1 | 0.0080 (2) | 0.0080 (2) | 0.0071 (3) | 0.0040 (1) | 0 | 0 |
| Sb1 | 0.0080 (2) | 0.0080 (2) | 0.0071 (3) | 0.0040 (1) | 0 | 0 |
| Ga2 | 0.0122 (2) | 0.0122 (2) | 0.0110 (2) | 0.0078 (2) | -0.0018(1) | 0.0018 (1) |
| Ga3 | 0.0088 (2) | 0.0147 (3) | 0.0143 (3) | 0.0073 (1) | 0.0001 (1) | 0.0003 (2) |
| Ga4 | 0.0104 (2) | 0.0104 (2) | 0.0095 (2) | 0.0046 (2) | -0.0004(1) | 0.0004 (1) |
| Ga5 | 0.0129 (2) | 0.0129 (2) | 0.0105 (2) | 0.0086 (2) | -0.0016(1) | 0.0016 (1) |
| Ga6 | 0.0142 (2) | 0.0142 (2) | 0.0090 (2) | 0.0079 (2) | 0.0019 (1) | -0.0019(1) |
| Ga7 | 0.0096 (3) | 0.0096 (3) | 0.0113 (4) | 0.0048 (1) | 0 | 0 |
| Cr1 | 0.0056 (3) | 0.0056 (3) | 0.0058 (5) | 0.0028 (2) | 0 | 0 |
| Cr2 | 0.0065 (3) | 0.0056 (2) | 0.0052 (3) | 0.0033 (2) | 0.0002 (3) | 0.0001 (1) |

$Cr_4PtGa_{16.45(1)}Sb_{0.55(1)}$:

| Atom | U11 | U22 | U33 | U12 | U13 | U23 |
|---|---|---|---|---|---|---|
| Pt1 | 0.0065 (1) | 0.0065 (1) | 0.0057 (2) | 0.0033 (1) | 0 | 0 |
| Ga1 | 0.0085 (4) | 0.0085 (4) | 0.0075 (5) | 0.0043 (2) | 0 | 0 |
| Sb1 | 0.0085 (4) | 0.0085 (4) | 0.0075 (5) | 0.0043 (2) | 0 | 0 |
| Ga2 | 0.0130 (3) | 0.0130 (3) | 0.0114 (4) | 0.0081 (3) | -0.0018(2) | 0.0018 (2) |
| Ga3 | 0.0132 (3) | 0.0132 (3) | 0.0110 (4) | 0.0086 (4) | -0.0016(2) | 0.0016 (2) |
| Ga4 | 0.0109 (3) | 0.0109 (3) | 0.0094 (4) | 0.0047 (3) | -0.0003(2) | 0.0003 (2) |
| Ga5 | 0.0092 (3) | 0.0154 (4) | 0.0145 (5) | 0.0077 (2) | 0.0004 (2) | 0.0007 (3) |
| Ga6 | 0.0146 (3) | 0.0146 (3) | 0.0089 (4) | 0.0081 (3) | 0.0019 (2) | -0.0019(2) |
| Ga7 | 0.0102 (4) | 0.0102 (4) | 0.0111 (7) | 0.0051 (2) | 0 | 0 |
| Cr1 | 0.0064 (4) | 0.0073 (5) | 0.0058 (5) | 0.0037 (2) | 0.0000 (2) | 0.0001 (4) |
| Cr2 | 0.0061 (5) | 0.0061 (5) | 0.0063 (10) | 0.0030 (3) | 0 | 0 |

$Cr_4PtGa_{16.46(1)}Sb_{0.56(1))}$:

| Atom | U11 | U22 | U33 | U12 | U13 | U23 |
|---|---|---|---|---|---|---|
| Pt1 | 0.0064 (1) | 0.0064 (1) | 0.0060 (1) | 0.0032 (1) | 0 | 0 |
| Ga1 | 0.0082 (2) | 0.0082 (2) | 0.0074 (3) | 0.0041 (1) | 0 | 0 |
| Sb1 | 0.0082 (2) | 0.0082 (2) | 0.0074 (3) | 0.0041 (1) | 0 | 0 |
| Ga2 | 0.0127 (2) | 0.0127 (2) | 0.0115 (2) | 0.0081 (2) | -0.0018(1) | 0.0018 (1) |
| Ga3 | 0.0129 (2) | 0.0129 (2) | 0.0110 (2) | 0.0083 (2) | -0.0017(1) | 0.0017 (1) |
| Ga4 | 0.0090 (2) | 0.0090 (2) | 0.0143 (3) | 0.0016 (2) | 0.0001 (1) | -0.0001() |
| Ga5 | 0.0142 (2) | 0.0142 (2) | 0.0094 (2) | 0.0079 (2) | 0.0019 (1) | -0.0019(1) |
| Ga6 | 0.0107 (2) | 0.0107 (2) | 0.0096 (2) | 0.0046 (2) | -0.0004(1) | 0.0004 (1) |
| Ga7 | 0.0099 (3) | 0.0099 (3) | 0.0117 (4) | 0.0050 (1) | 0 | 0 |
| Cr1 | 0.0060 (2) | 0.0060 (2) | 0.0057 (3) | 0.0025 (3) | -0.0001(1) | 0.0001 (1) |
| Cr2 | 0.0060 (3) | 0.0060 (3) | 0.0063 (5) | 0.0030 (2) | 0 | 0 |

**Table S5.** Fitting parameters of $C_p/T$ vs $T^2$ curves via $C_p/T = \beta_1 T^2 + \beta_2 T^4 + \beta_3 T^6 + \beta_4 T^8 + \beta_5 T^{10}$ for CPGS-2.94%, CPGS-3.24% and CPGS-3.29%.

| | **β1** | **β2** | **β3** | **β4** | **β5** |
|---|---|---|---|---|---|
| **CPGS-2.94%** | 0.0022 (1) | $-1.1(1)\times10^{-5}$ | $1.7(2)\times10^{-7}$ | $-5.4(5)\times10^{-10}$ | $5.5(5)\times10^{-13}$ |
| **CPGS-3.24%** | 0.0015 (1) | $-0.5(1)\times10^{-5}$ | $0.9(1)\times10^{-7}$ | $-3.0(3)\times10^{-10}$ | $3.1(2)\times10^{-13}$ |
| **CPGS-3.29%** | 0.0013 (1) | $-0.5(2)\times10^{-5}$ | $0.9(1)\times10^{-7}$ | $-2.9(3)\times10^{-10}$ | $2.9(3)\times10^{-13}$ |

**Figure S1.** Proposed magnetic entropy change ($\Delta S_{mag}$) of CPGS-2.94%, CPGS-3.24% and CPGS-3.29%.

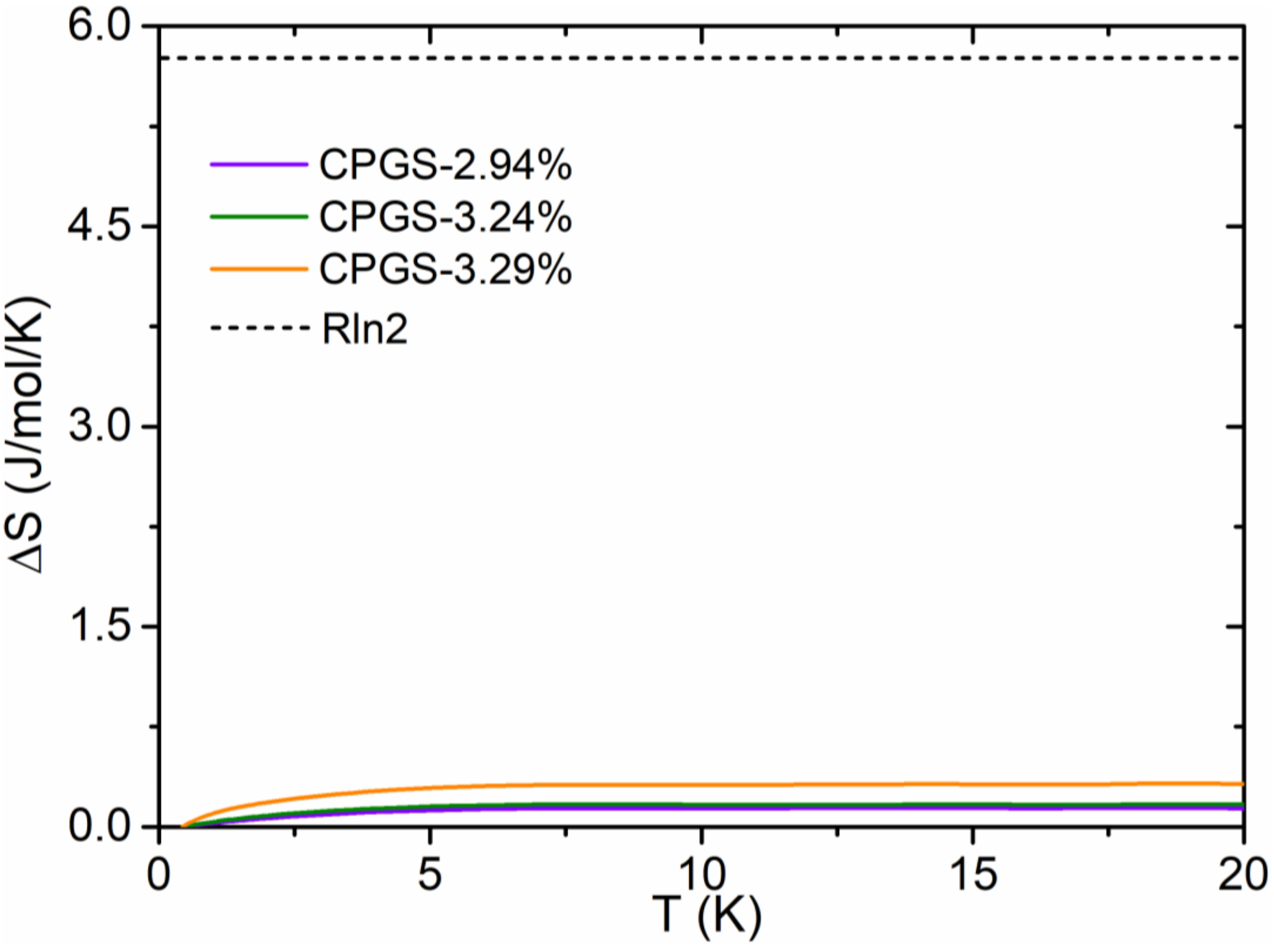